# Study on Asymmetric Diffraction of Acoustic Parity-Time-Symmetric Gratings Using Rigorous Coupled-Wave Analysis

Yuzhen Yang,[1, 2] Han Jia,[1, 2, 3,*] Jun Yang[1, 2, 3, †]

[1] Key Laboratory of Noise and Vibration Research, Institute of Acoustics, Chinese Academy of Sciences, Beijing 100190, People's Republic of China

[2] University of Chinese Academy of Sciences, Beijing 100049, People's Republic of China

[3] State Key Laboratory of Acoustics, Institute of Acoustics, Chinese Academy of Sciences, Beijing 100190, People's Republic of China

In *PT*-symmetric gratings, asymmetric diffraction can be generated by modulating the ratio between imaginary and real refractive indices. In this paper, a rigorous coupled-wave analysis (RCWA) has been developed to analyze the diffraction properties of acoustic *PT*-symmetric gratings with two kinds of modulating approaches, including modulating the effective modulus and the effective density. Asymmetric diffraction with both Bragg incident angles and perpendicular incident angles is discussed by using the RCWA method. Results show that the modulation ratio for the diffraction vanishing point changes with the modulation amplitude differently for two kinds of modulating approaches. Moreover, the sound energy will be weaken or be enhanced at Bragg incident angles depending on the sign of incident angles and the modulation ratio.

*hjia@mail.ioa.ac.cn

†jyang@mail.ioa.ac.cn

## I. INTRODUCTION

Parity-time (*PT*) symmetry is a concept arising from quantum mechanics to describe the invariance of Hamiltonian under the combination of parity (*P*) and time-reversal (*T*) operations [1]. Because of the similarity between Schrödinger equation and classic wave equation, non-Hermitian systems with *PT*-symmetric potential have been widely discussed in fields of classic waves [2-30]. In the *PT*-symmetric systems of classic waves, distribution of complex refractive index needs to be a Hermitian function of position as $n(x) = n^*(-x)$, where the superscript $*$ denotes the conjugation. This distribution requires precisely balancing loss and gain materials. Many interesting phenomena in *PT*-symmetric systems of classic waves have been explored, such as one-way cloak [16-18], unidirectional reflection [20-22] and coherent perfect absorption and lasing [10-13, 24, 29]. These novel effects derive from the phase transition point of *PT*-symmetry, where the system Hamiltonian switches between real and complex spectrum. Recent progress on *PT*-symmetric systems yields new schemes to realize asymmetric transport without nonlinear effect or mode conversion [16-23, 27]. Such as in *PT*-symmetric diffraction gratings, asymmetric diffraction is observed between a pair of oblique incidence waves at the exceptional point (EP) [23, 30, 31]. Compared to conventional diffraction gratings that have symmetric diffraction orders, *PT*-symmetric gratings exhibit intrinsically asymmetric diffraction property that the forward diffracted positive first order disappears completely when the real and imaginary parts of refractive index are equal.

In previous researches, two-wave coupled theory is always used to analyze the sound field in *PT*-symmetric gratings, which neglects the boundary diffraction and the second derivatives of the field amplitudes in gratings [21-24, 29-31]. This approximation is convenient when the modulation is assumed to be very small. However, higher orders of space harmonics will become more and more important with increasing the modulation amplitude so that the approximation will cause large errors in calculation. A rigorous analysis model has been presented to analyze acoustic gratings made of perfect rigid bodies [32-34]. In their model, the pressure field inside a single aperture is assumed as the superposition of counter propagating waveguide modes, which only can be applied to gratings with air apertures surrounded by rigid walls. In this paper, we presented a rigorous coupled-wave analysis (RCWA) to study the diffraction properties of acoustic *PT*-symmetric diffraction gratings. This method is developed from the solution of electromagnetic wave diffraction in planar gratings [35-43], which is a straightforward solution based on the governing set of differential equations. In this method, the pressure field inside a grating is considered as the composition of numbers of plane waves, including forward-traveling waves and backward-traveling waves. The diffraction efficiencies are determined by solving the acoustic wave equation in the grating and matching the continuity conditions at boundaries of grating.

By using our proposed RCWA method in acoustics, we have studied the diffraction properties of acoustic *PT*-symmetric diffraction gratings. The diffraction vanishing point (DVP) corresponding to forward diffracted positive first order disappeared and the sound energy variation in gratings will be discussed as main objects. The modulation ratio between real and imaginary parts for the DVP changes with the modulation amplitude. This phenomenon is different from the results of previous researches that complete diffraction suppression of *PT*-symmetric gratings always occurs when the amplitudes of the real and imaginary parts are identical [23, 30, 31]. Moreover, the sound energy will be weaken or be enhanced in the diffraction gratings depending on both the sign of incident angles and the modulation ratio.

This paper is organized as follows. After this introduction, Sec. II describes both the theory and solution of RCWA in acoustic diffraction gratings. Afterwards, Sec. III presents a detailed discussion on two-beam diffraction of *PT*-symmetric gratings at Bragg incident angles. Two kinds of approaches to modulate the distribution of refractive index are presented, including modulating the effective modulus and the effective density. Full-wave simulations are also given to validate the developed RCWA method. Section IV studies the multi-beam diffraction properties with perpendicular incidence. Then Sec. V discusses the convergence of RCWA briefly, for the sake of completeness. Finally, Sec. VI summarizes the main findings of this work.

## II. RIGOROUS COUPLED-WAVE ANALYSIS OF ACOUSTIC DIFFRACTION GRATING

The acoustic diffraction grating is schematically illustrated in Fig. 1. The grating fringes are periodical in *x* direction. When acoustic waves transmit the grating, both backward and forward diffracted waves will be produced. To conveniently describe the diffraction phenomenon, we divide the whole area into three regions: regions I and III are homogeneous air background; region II represents the area within the modulated grating. The normalized pressure in three areas can be expressed as follows:

$$P_1 = \exp(-jk_{x0}x - jk_{z0}z) + \sum_i R_i \exp(-jk_{xi}x + jk_{zi}z), \tag{1}$$

$$P_2 = \sum_i S_i(z)\exp(-jk_{xi}x), \tag{2}$$

$$P_3 = \sum_i T_i \exp[-jk_{xi}x - jk_{zi}(z-d)], \tag{3}$$

where $k_{x0} = k_0 \sin\theta$, $k_{z0} = k_0 \cos\theta$, $k_{xi} = k_0 \sin\theta + iK$ ($K = 2\pi/T$) for any integer $i$ (the wave index), $k_{zi} = +\sqrt{k_0^2 - k_{xi}^2}$ for $k_{xi} < k_0$ and $k_{zi} = -j\sqrt{k_{xi}^2 - k_0^2}$ for $k_{xi} > k_0$. Here $k_0$ is the wave vector in air; $\theta$ is the incident angle; $T$ and $d$ are the period and thickness of the grating; $R_i$ and $T_i$ are normalized amplitude of the *i*th reflected wave and transmitted wave, respectively; $S_i$ is the *i*th wave field at any point within the modulated grating. For a given index of *i*, the wave field inside the grating can be expressed as a superposition of an infinite number of plane waves. This superposition includes forward-traveling waves and backward-traveling waves. These amplitudes are determined by solving the acoustic wave equation $\nabla^2 P + k_0^2 n^2(r)P = (\nabla \ln \rho(r)) \cdot \nabla P$ for inhomogeneous modulated-region, which can be simplified in this *x*-direction periodic acoustic grating as

$$\nabla^2 P + k_0^2 n^2(x)P - \frac{\partial \ln \rho(x)}{\partial x}\frac{\partial P}{\partial x} = 0, \tag{4}$$

where $n(x)$ and $\rho(x)$ are position dependent refractive index and mass density in

the modulated grating. Because of the periodicity, $n^2(x)$ and $\ln(\rho(x))$ can be expanded as follows:

$$n^2(x) = \sum_m x_m \exp(-jmKx),$$
$$\ln \rho(x) = \sum_m y_m \exp(-jmKx). \tag{5}$$

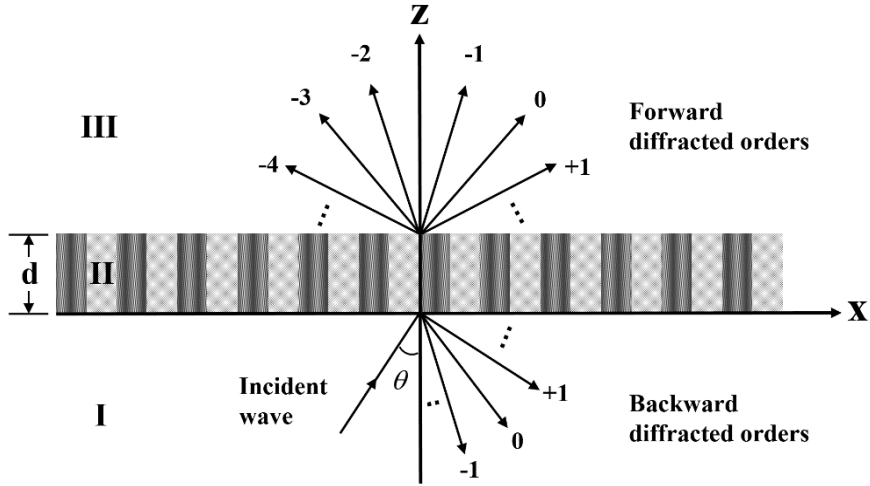

FIG. 1. Geometry for an acoustic diffraction grating.

To solve the wave field $S_i(z)$ in the grating, we substitute Eqs. (2) and (5) into acoustic wave equation [Eq.(4)] and obtain the rigorous coupled-wave equation:

$$S_i''(z) - k_{xi}^2 S_i(z) + k_0^2 \sum_p x_{(i-p)} S_p(z) + \sum_q (i-q) K k_{xq} y_{(i-q)} S_q(z) = 0. \tag{6}$$

Equation (6) could be written in matrix form as $[S''] = [A][S]$, where $S$ and $S''$ are the column vectors of $S_i$ and $d^2 S_i / dz^2$, respectively. The expression of matrix $A$ is $A = K_x^2 - k_0^2 X - K K_x Y$ ; $X$, $Y$ are both matrixes, whose elements are $X_{m,n} = x_{m-n}$, $Y_{m,n} = (m-n) y_{m-n}$ ,respectively; $K_x$ is a diagonal matrix, whose elements are $K_{i,i} = k_{xi}$. The system given by differential equations [Eq. (6)] has a simple and straightforward solution, which can be obtained in terms of the eigenvalues and the eigenvectors of the matrix $A$. It is

$$S_i(z) = \sum_m w_{i,m} \{c_m^+ \exp(-q_m z) + c_m^- \exp[q_m(z-d)]\}, \tag{7}$$

where $q_m$ is the square root of the $m$th eigenvalue and $w_{i,m}$ is the $m$th element of the row in the eigenvector matrix $\mathbf{W}$. The coefficients $c_m^+$ and $c_m^-$ can be determined by matching the sound pressure and normal particle vibration velocity at the boundaries $z=0$ and $z=d$. The normal particle vibration velocity can be expressed as $v_z = -\dfrac{1}{j\omega\rho}\dfrac{\partial P}{\partial z}$ for simple harmonic wave. By expanding the normalized reciprocal of mass density as $\rho_0 \dfrac{1}{\rho} = \sum_m z_m \exp(-jmKx)$, where $\rho_0$ is the mass density of air, the $i$th particle vibration velocity can be expressed as follows:

$$v_{z,i} = -\frac{1}{j\omega\rho_0}\sum_l z_{(i-l)}\sum_m w_{l,m}\{-q_m c_m^+ \exp(-q_m z) + q_m c_m^- \exp[q_m(z-d)]\}. \tag{8}$$

By combining Eqs (1), (2), (3), (7) and (8), the continuity conditions at $z=0$ can be expressed as:

$$\begin{aligned}\delta_{i0} + R_i &= \sum_m w_{i,m}[c_m^+ + c_m^- \exp(-q_m d)], \\ -jk_0(\cos\theta)\delta_{i0} + jk_{zi}R_i &= \sum_l z_{(i-l)}\sum_m w_{l,m}[-q_m c_m^+ + q_m c_m^- \exp(-q_m d)]. \end{aligned} \tag{9}$$

The continuity conditions at $z=d$ are shown as:

$$\begin{aligned}T_i &= \sum_m w_{i,m}[c_m^+ \exp(-q_m d) + c_m^-], \\ -jk_{zi}T_i &= \sum_l z_{(i-l)}\sum_m w_{l,m}[-q_m c_m^+ \exp(-q_m d) + q_m c_m^-]. \end{aligned} \tag{10}$$

Here $\delta_{i0}$ is the Kronecker delta function. The boundary conditions can be written as matrix form:

$$\begin{bmatrix}\mathbf{\Delta}_{i0} \\ jk_0(\cos\theta)\mathbf{\Delta}_{i0}\end{bmatrix} + \begin{bmatrix}\mathbf{I} \\ -j\mathbf{K_z}\end{bmatrix}[\mathbf{R}] = \begin{bmatrix}\mathbf{W} & \mathbf{WE} \\ \mathbf{ZWQ} & -\mathbf{ZWQE}\end{bmatrix}\begin{bmatrix}\mathbf{C^+} \\ \mathbf{C^-}\end{bmatrix}, \tag{11}$$

$$\begin{bmatrix}\mathbf{WE} & \mathbf{W} \\ \mathbf{ZWQE} & -\mathbf{ZWQ}\end{bmatrix}\begin{bmatrix}\mathbf{C^+} \\ \mathbf{C^-}\end{bmatrix} = \begin{bmatrix}\mathbf{I} \\ j\mathbf{K_z}\end{bmatrix}[\mathbf{T}], \tag{12}$$

where $\mathbf{R}$, $\mathbf{T}$ are column vectors of reflected waves and transmitted waves; $\mathbf{\Delta}_{i0}$ is a column vector, in which only the element corresponding to zero-order wave equals to 1 and others are zero; $\mathbf{Z}$ is a matrix, whose elements are $Z_{m,n} = z_{m-n}$; $\mathbf{Q}$, $\mathbf{E}$, $\mathbf{K_z}$ are diagonal matrixes, whose elements are $q_m$, $\exp(-q_m d)$, $k_{zi}$, respectively; $\mathbf{C}^+$, $\mathbf{C}^-$ are column vectors of coefficients $c_m^+$, $c_m^-$. Using matrix algebra to eliminate $\mathbf{R}$ and $\mathbf{T}$ in Eqs. (11) and (12) gives

$$\begin{bmatrix} j\mathbf{K}_z\Delta_{i0} + jk_0(\cos\theta)\Delta_{i0} \\ 0 \end{bmatrix} = \begin{bmatrix} j\mathbf{K}_z\mathbf{W}+\mathbf{V} & j\mathbf{K}_z\mathbf{WE}-\mathbf{VE} \\ j\mathbf{K}_z\mathbf{WE}-\mathbf{VE} & j\mathbf{K}_z\mathbf{W}+\mathbf{V} \end{bmatrix}\begin{bmatrix} \mathbf{C}^+ \\ \mathbf{C}^- \end{bmatrix}, \quad (13)$$

where $\mathbf{V}=\mathbf{ZWQ}$ for the sake of simplicity. The column vectors of coefficients can be solved from Eq. (13), and then column vectors of reflected waves and transmitted waves can be obtained as $[\mathbf{R}] = [\mathbf{W} \quad \mathbf{WE}]\begin{bmatrix} \mathbf{C}^+ \\ \mathbf{C}^- \end{bmatrix} - \Delta_{i0}$ and $[\mathbf{T}] = [\mathbf{WE} \quad \mathbf{W}]\begin{bmatrix} \mathbf{C}^+ \\ \mathbf{C}^- \end{bmatrix}$. The diffraction efficiencies for different diffracted orders, which is the ratio between energy of diffracted order and energy of incident wave, are given by $DE_{1i} = R_i R_i^* \text{Re}(k_{zi}/k_{z0})$, $DE_{3i} = T_i T_i^* \text{Re}(k_{zi}/k_{z0})$, where $DE_{1i}$ and $DE_{3i}$ are the diffraction efficiencies in regions I and III.

## III. TWO-BEAM DIFFRACTION OF *PT*-SYMMTRIC GRATINGS AT BRAGG INCIDENT ANGLE

Figure 2 shows a *PT*-symmetric grating, in which the real modulated refractive index ($n_r$) is an even function of position and the imaginary modulated refractive index ($n_i$) is an odd function of position. One period of the grating that is indicated by the grey rectangular box contains four parts with equal duty cycle. The enlarged view is shown in the right inset, in which four parts are marked by 1, 2, 3 and 4. Refractive indices of different parts are expressed as $n_1 = (1+n')+i\delta n'$, $n_2 = (1-n')+i\delta n'$, $n_3 = (1-n')-i\delta n'$, $n_4 = (1+n')-i\delta n'$, respectively. $n'$ and $\delta$ are the modulation amplitude of real refractive index and the amplitude ratio between imaginary modulated part and real modulated part. The positive imaginary refractive indices ($n_i > 0$) of parts 1 and 2 represent gain and the negative imaginary refractive indices ($n_i < 0$) of parts 3 and 4 represent loss. The propagation of acoustic waves is generally manipulated by two kinds of methods: modulating the effective modulus or the effective density of unit cells. So gratings modulated by two different approaches are discussed in this paper.

We firstly discussed the two-beam diffraction of the *PT*-symmetric gratings with Bragg incident angles. The period of grating is set equal to the wave length, that is, $T=\lambda$. In this situation, only zero-order and first-order diffracted waves are constructed at Bragg incident angle ($\theta_B = 30°$). All higher orders are evanescent waves since the corresponding horizontal wave vectors are larger than $k_0$. In this situation, the angles between two forward and two backward diffracted orders are both $60°$, and it is easy to recognize different diffracted orders in full wave simulation with a finite element (FEM) software *Comsol Multiphysics*. So full-wave simulations of spatial Gaussian incident beams are also reported to prove the correctness of the developed RCWA method.

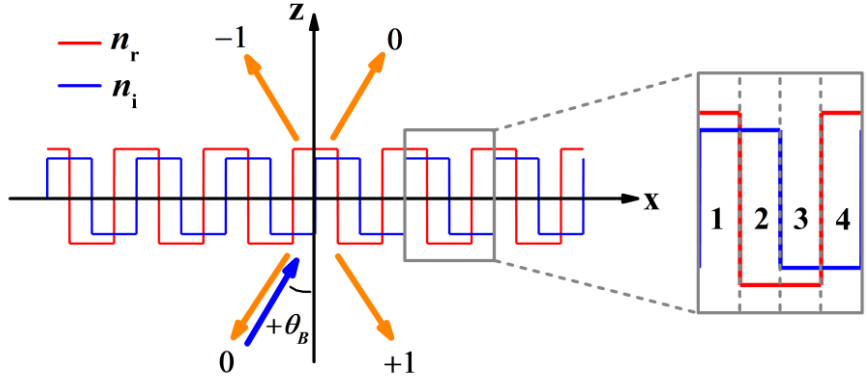

FIG. 2. Schematic diagram of diffraction at Bragg incident angle in an acoustic *PT*-symmetric grating. $n_r$ ($n_r = \pm n'$) and $n_i$ ($n_i = \pm \delta n'$) are the real and imaginary parts of modulated refractive index. Blue arrow represents incident wave and orange arrows represent corresponding diffracted waves. Inset is enlarged view of the period in grey rectangular box.

### A. Modulating the effective modulus

In the case of modulating the effective modulus, the mass density of grating is set as homogenous and equals to the mass density of air $\rho_0$. Figure 3 shows the diffraction efficiencies of *PT*-symmetric grating when the modulation amplitude $n'$ equals to 0.05. Figure 3(a) shows the total diffraction efficiencies with positive and negative Bragg incident angles. Total diffraction efficiency is the sum of diffraction efficiencies of all forward and backward diffracted orders. When the incident angle is positive, the total diffraction efficiency is larger than 1, which indicates that gain plays a leading role in the propagation process. When the incident angle is negative, the total diffraction efficiency is lower than 1 before the DVP and higher than 1 after the DVP. The DVP transforms an absorbing grating into an amplifying grating for the case of negative incident angle. Then diffraction efficiencies of specific diffracted orders are discussed. The backward diffracted orders are weak because the modulations of gratings are relatively small. In this case, the diffraction efficiencies of forward diffracted orders are studied. As shown in Fig. 3(b), diffraction efficiencies of forward zero orders with both positive and negative incident angles are equal. However, the diffraction efficiencies of forward first orders (Figs. 3(c) and (d)) are quite different when the incident angle changes from positive to negative. The positive first-order diffracted wave with negative incident angle is suppressed completely when modulation ratio $\delta$ equals to 0.99, which is very close to previous results of two-wave coupled theory [23, 30, 31]. The amplitudes of pressure fields in one period at $\delta=0.99$ are shown in the insets of Figs. 3(c) and (d). For positive incident angle, the higher sound pressure is mainly localized in gain parts, while the sound pressure is almost evenly distributed at gain and loss parts for negative incident angle. The amplitudes of pressure fields well explain why the total diffraction efficiency is higher than 1 with positive incident angle but is equal to 1 with negative incident angle at the DVP. All above results shown in Fig. 3 are obtained by RCWA (lines) and FEM (symbols). It can be observed that the results of numerical calculation by RCWA agree well with the simulation results by FEM.

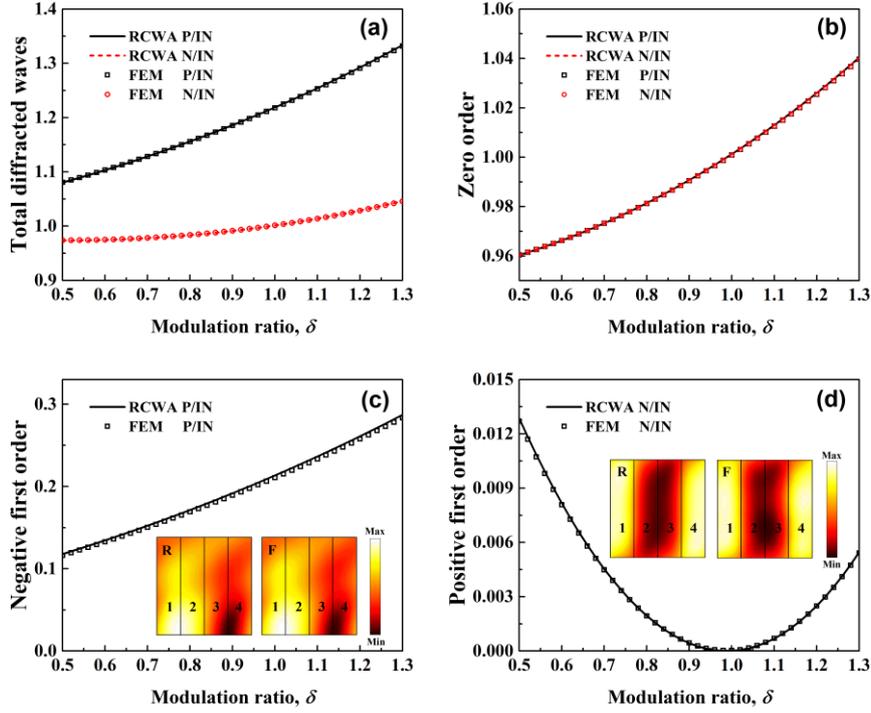

FIG. 3. Diffraction efficiencies for the case of modulating the effective modulus with modulation amplitude $n'=0.05$. RCWA and FEM represent the results of numerical calculation by RCWA and simulation results by FEM; P/IN and N/IN represent cases of positive and negative incident angles. Diffraction efficiencies for (a) total diffracted waves, (b) forward diffracted zero order, (c) forward diffracted negative first order, (d) forward diffracted positive first order. Insets: amplitudes of pressure fields in one period at the DVP obtained by RCWA (R) and FEM (F).

In order to enhance the diffraction capability, the modulation amplitude is increased to $n'=0.2$. The diffraction efficiencies are shown in Fig. 4. All the diffraction efficiencies are greatly improved after increasing the modulation amplitude. In terms of the variation tendency of diffraction efficiencies with the modulation ratio, the diffraction properties are almost the same as the diffraction of above grating with $n'=0.05$. When the incident angle is positive, sound energy is amplified a lot in the diffraction grating. For negative incident angle, the diffraction grating also changes from an absorber to an amplifier at the DVP. The amplitudes of pressure fields at the DVP are shown in insets of Figs. 4(c) and (d). Sound energy is also mainly localized in gain parts for positive incident angle while evenly distributed in gain and loss parts for negative incident angle. Asymmetric diffraction still exists between positive and negative incident angles. However, it is worth noting that the DVP does not locate near to $\delta=1$ any longer but moves forward to $\delta=0.86$. Previously, the DVP always occurs when the imaginary modulation equals to the real modulation in *PT*-symmetric diffraction gratings. This result is obtained using two-wave coupled theory and neglecting backward diffracted orders, which is only applicable when the modulation is very small. After increasing the modulation amplitude, backward and higher diffracted orders become influential and cannot be neglected. So the two-wave coupled theory is not accurate any longer. However, as a rigorous solution, the proposed RCWA, which contains every forward and backward diffracted order, is still applicable in the case of large modulation amplitude. Theoretically, this rigorous and straightforward solution is useable in all applications of planar gratings. The results of RCWA (lines in Fig. 4) are still consistent with the results of FEM (symbols in Fig. 4), which is a proof for the accuracy of RCWA.

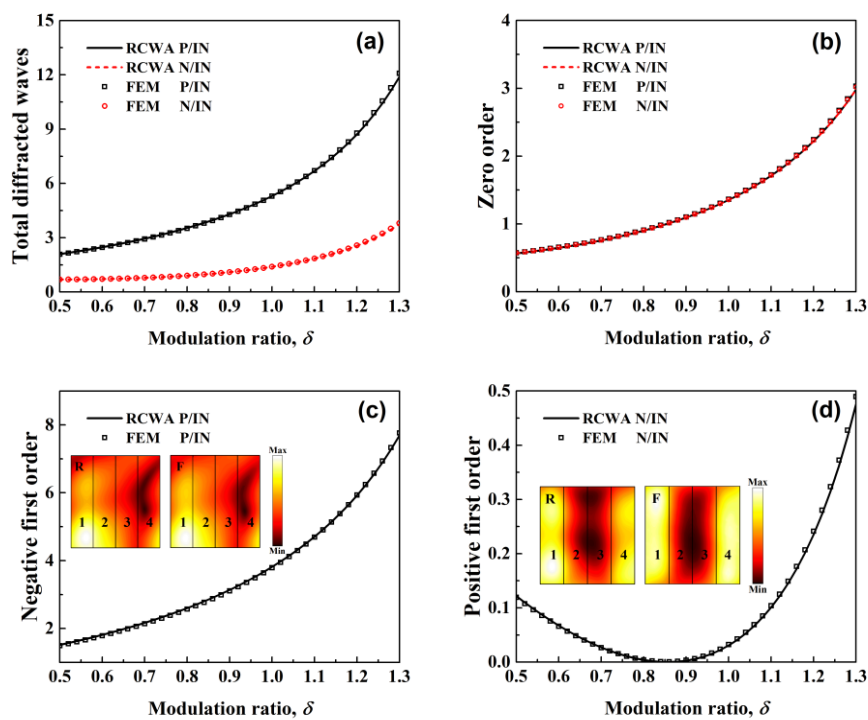

FIG. 4. Diffraction efficiencies for the case of modulating the effective modulus with modulation amplitude $n'=0.2$. Diffraction efficiencies for (a) total diffracted waves, (b) forward diffracted zero order, (c) forward diffracted negative first order, (d) forward diffracted positive first order. Insets: amplitudes of pressure fields in one period at the DVP.

## B. Modulating the effective density

In the case of modulating the effective density, the effective densities of different parts are set as $\rho_l = \text{Re}(n_l)^2 \rho_0$ ($l=1,2,3,4$), which are proportional to the square of real refractive indices. All the other conditions are the same as the case of modulating the effective modulus. Figures 5 and 6 show the diffraction efficiencies with modulation amplitudes $n'=0.05$ and $n'=0.2$, respectively. The diffraction properties of grating modulating the effective density are almost the same as grating modulating the effective modulus. Sound energy also increases in the diffraction grating when the incident angle is positive, and changes from reductive to amplifying across the DVP when the incident angle is negative. As the amplitudes of pressure fields shown (insets of Figs. 5 and 6), sound energy is still localized in gain parts for positive incident angle while evenly distributed in gain and loss parts for negative incident angle. However, the DVPs of modulating the effective density are quite different from the DVPs of modulating the effective modulus. The modulation ratio of the DVP changes from $\delta=0.99$ to $\delta=0.50$ when the modulation amplitude $n'$ equals to 0.05, and the modulation ratio of DVP changes from $\delta=0.86$ to $\delta=0.56$ when the modulation amplitude $n'$ equals to 0.2.

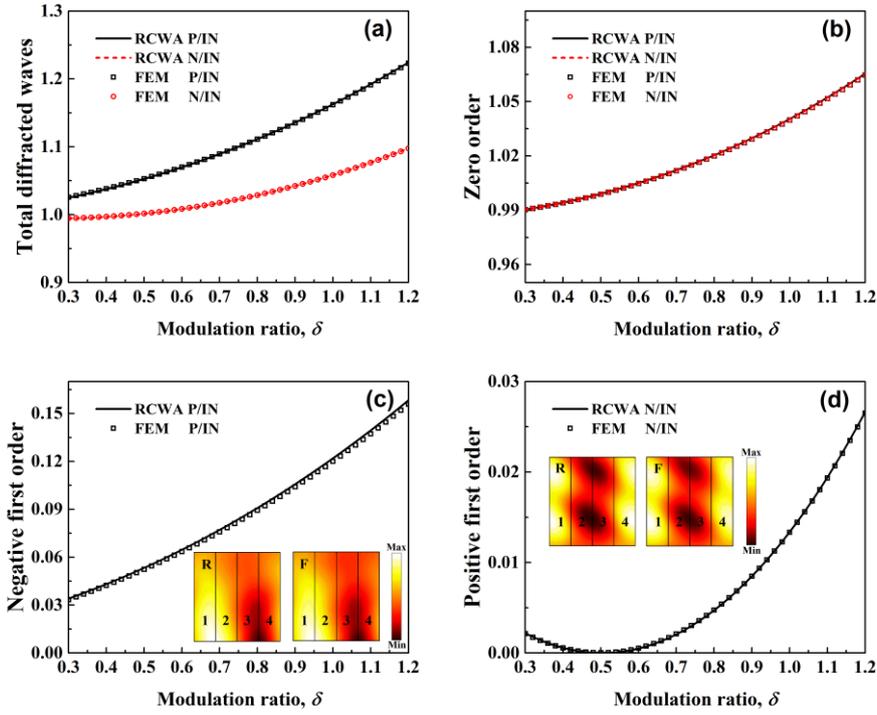

FIG. 5. Diffraction efficiencies for the case of modulating the effective density with modulation amplitude $n'$ equals to 0.05. Diffraction efficiencies for (a) total diffracted waves, (b) forward diffracted zero order, (c) forward diffracted negative first order, (d) forward diffracted positive first order. Insets: amplitudes of pressure fields in one period at the DVP.

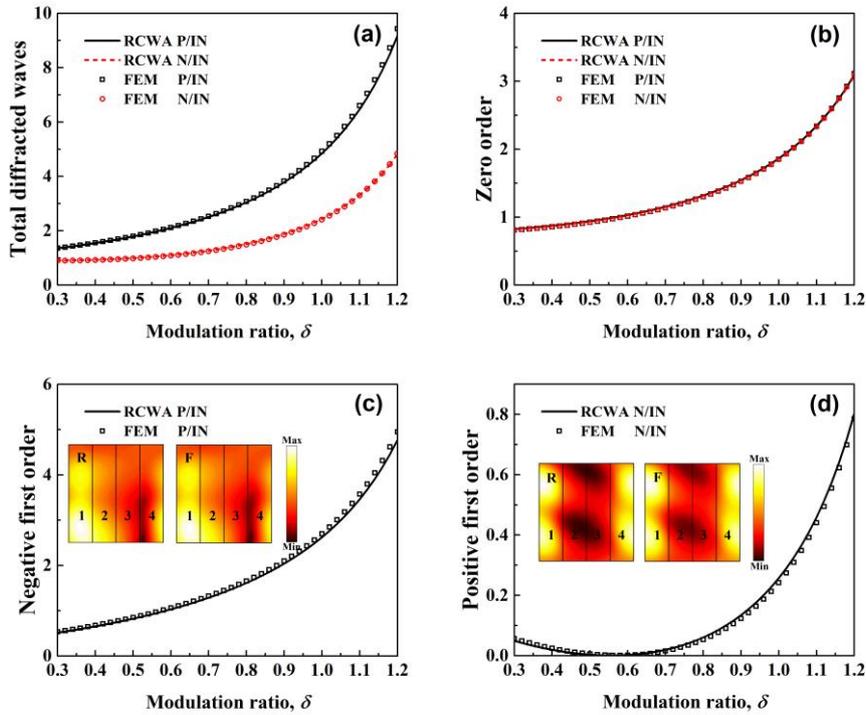

FIG. 6. Diffraction efficiencies for the case of modulating the effective density with modulation

amplitude $n'$ equals to 0.2. Diffraction efficiencies for (a) total diffracted waves, (b) forward diffracted zero order, (c) forward diffracted negative first order, (d) forward diffracted positive first order. Insets: amplitudes of pressure fields in one period at the DVP.

### C. Variation of the DVP at Bragg incident angle

From the above results, it is clear that both the modulation amplitudes and modulating approaches will influence the modulation ratio for the DVP. The detailed variations of the DVP for two modulating approaches ($\rho = \rho_0$ and $\rho = \text{Re}(n)^2 \rho_0$) are calculated by both RCWA and FEM. Figure 7 shows that with increasing the modulation amplitude, the modulation ratio of the DVP decreases for the case of modulating the effective modulus but increases for the case of modulating the effective density. When the modulation amplitude is small, the difference of densities between two kinds of modulating approaches is also small. However, the modulation ratios for the DVPs are counterintuitively different. In fact, both the inner diffraction and the boundary diffraction will influence the energy distribution for every diffraction orders. From the theory and solution of RCWA, we can see that the refractive index distribution only affects the inner diffraction, but the density distribution affects both the inner diffraction and the boundary diffraction. When the boundary diffraction plays an important role in the diffraction progress, a small difference of density modulation could lead a large difference between the modulation ratios for the DVPs.

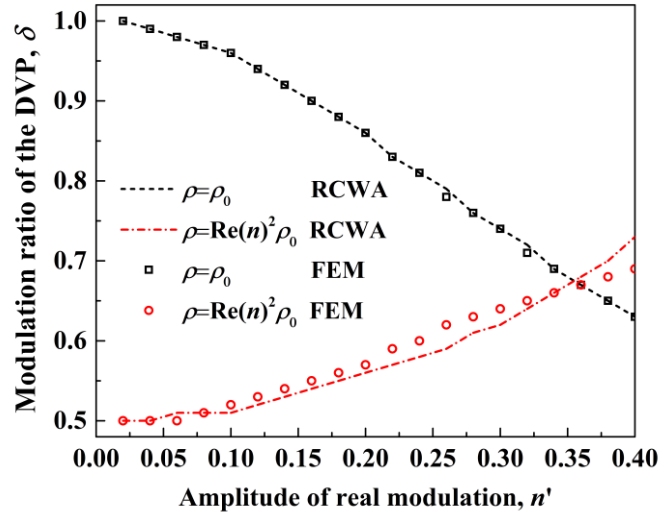

FIG. 7. Relationships between the modulation ratio for the DVP and the modulation amplitude at Bragg incident angle. $\rho = \rho_0$ represents the case of modulating the effective modulus and $\rho = \text{Re}(n)^2 \rho_0$ represents the case of modulating the effective density.

## IV. MULTI-BEAM DIFFRACTION WITH PERPENDICULAR INCIDENCE

This part will discuss a more general situation: diffraction properties of *PT*-symmetric grating with higher diffracted orders. Sound waves are perpendicularly incident into a grating with a period equaling to five times of wave length, that is, $T = 5\lambda$. Then multiple diffracted beams will be generated in forward and backward diffracted regions. Figure 8 shows the diffraction efficiencies of first-order diffracted waves with different modulation amplitudes for both modulating approaches. The diffraction efficiencies of forward diffracted positive first order are much lower than those of negative first order. In the case of perpendicular incidence, the diffraction efficiencies for first-order diffracted waves with modulating the effective modulus and the effective density are similar to each other. As increasing the modulation amplitude and modulation ratio, small difference between two modulating approaches will appear. Figure 9 shows the relationships between modulation ratio for the DVP and the modulation amplitude, which are nearly identical for the case of modulating the effective modulus and the case of modulating the effective density. The modulation ratio for the DVP increases from $\delta=0.98$ to $\delta=1.38$ with increasing the modulation amplitude, which is quite different from the case of Bragg incident angle.

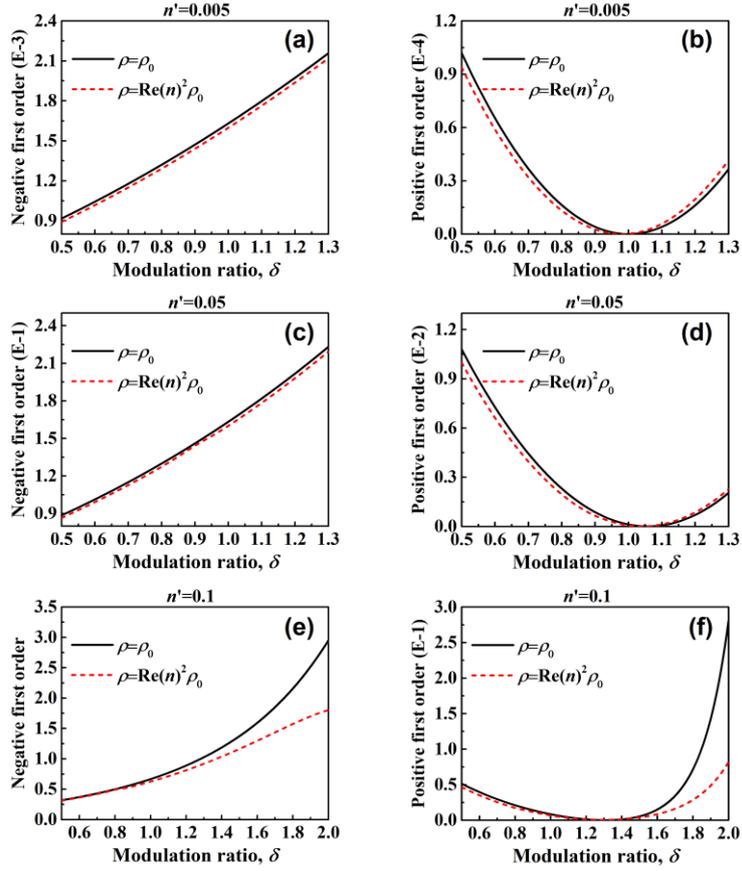

FIG. 8. Diffraction efficiencies of first-order diffracted waves with different modulation amplitudes for two kinds of modulating approaches. Diffraction efficiencies for negative first order with (a) $n'=0.005$, (c) $n'=0.05$, (e) $n'=0.1$, and for positive first order with (b) $n'=0.005$, (d) $n'=0.05$, (f) $n'=0.1$.

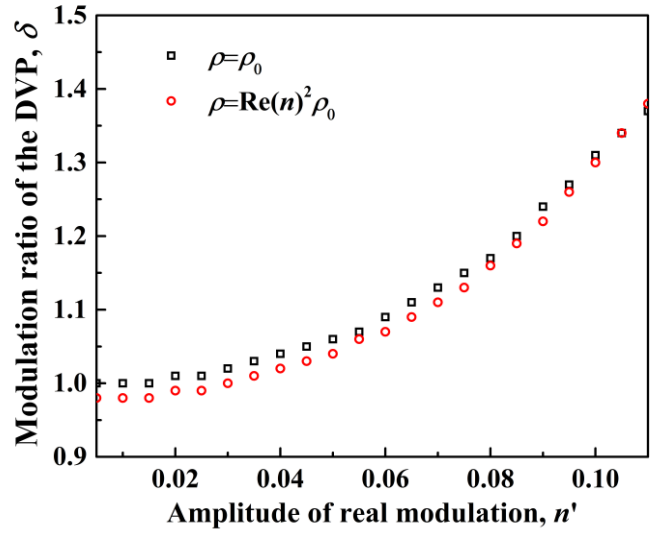

FIG. 9. Relationships between the modulation ratio for the DVP and the modulation amplitude with perpendicular incidence.

## V. NUMERICAL STABILITY

The criteria of numerical stability for Hermitian gratings are energy conservation and convergence to the proper solution with an increasing number of space harmonics. The gratings in this paper are non-Hermitian systems possessing $PT$-symmetric refractive index $n(x) = n^*(-x)$, so the energy conservation is not under consideration. We only discussed the convergence of RCWA with an increasing number of space harmonics. Without loss of generality, we randomly selected relative parameters for the diffraction gratings. The period of grating $T$ equals to $8\lambda$ and the incident angle $\theta$ is $20^o$. In Fig. 10, convergence of two modulating methods ( $\rho = \rho_0$ and $\rho = \text{Re}(n)^2 \rho_0$ ) and two modulation amplitudes ( $n' = 0.05$ and $n' = 0.2$ ) are exhibited. It is clear that the diffraction efficiencies converge to the proper values when a sufficient number of space harmonics are included in the formulation. More harmonics are required for gratings with larger modulation amplitude.

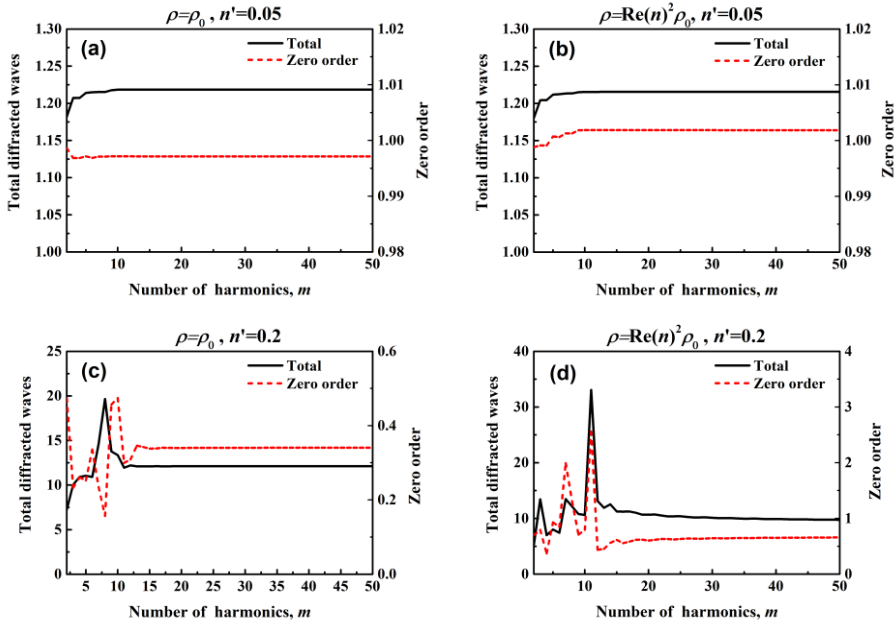

FIG. 10. Convergence of diffraction efficiencies dependence on the number of space harmonics for two kinds of modulating approaches and two kinds of modulation amplitudes: modulating the modulus with (a) $n'=0.05$, (c) $n'=0.2$, and modulating the density with (b) $n'=0.05$, (d) $n'=0.2$.

## VI. CONCLUSION AND DISCUSSION

In this paper, we have developed the RCWA to calculate the diffraction of acoustic planar gratings. By using the proposed RCWA, we have calculated and analyzed the diffraction properties of acoustic *PT*-symmetric gratings for two-beam diffraction at Bragg incident angles and multi-beam diffraction with perpendicular incidence. At Bragg incident angles, asymmetric diffraction occurs when the incident angle changes from positive to negative symmetrically. The forward diffracted positive first order with negative incident angle disappears completely at the DVP. The modulation ratio for the DVP decreases with the increasing of the modulation amplitude for the case of modulating the effective modulus, while the modulation ratio for the DVP increases with the increasing of the modulation amplitude for the case of modulating the effective density. For multi-beam diffraction with perpendicular incidence, asymmetric diffraction still can be observed between symmetric diffracted orders. The diffraction efficiencies of forward diffracted positive first order are much lower than those of negative first order. The modulation ratio for the DVP increases with the increasing of the modulation amplitude for both modulating approaches. And small difference between two modulating approaches will appear with the increasing of the modulation amplitude and modulation ratio.

In two-beam diffraction at Bragg incident angle, the sound energy in the diffraction grating for positive incident angle is much larger than the sound energy for negative incident angle. From the amplitudes of pressure fields at the DVP, it can be seen that higher sound pressure is mainly localized in gain parts for positive incident angle while sound pressure is almost evenly distributed in gain and loss parts for negative incident angle. Full-wave simulations are also reported and the results of numerical calculation by RCWA are in good agreement with the simulation results by FEM, which support a sufficient proof for the accuracy of RCWA.

The solution of RCWA can be easily implemented by a digital computer in a simple matrix form, which is faster and more efficient than FEM in the application of planar gratings. Especially for multi-beam diffraction, RCWA calculated the diffraction efficiency of every diffracted order directly. But it is difficult to distinguish different diffracted orders straightly in the real space in the simulation of FEM.

## ACKNOWLEDGEMENT

Discussions and technical assistance from Dr. Jiaheng Li are gratefully acknowledged. This work is supported by the National Natural Science Foundation of China (Grant No. 11874383), the Youth Innovation Promotion Association CAS (Grant No. 2017029), and the IACAS Young Elite Researcher Project (Grant No. QNYC201719).

**References**

[1] M. Bender and S. Boettcher, Real Spectra in Non-Hermitian Hamiltonians Having *PT* Symmetry, Phys. Rev. Lett. **80**, 5243 (1998).

[2] K. G. Makris, R. El-Ganainy, D. N. Christodoulides and Z. H. Musslimani, Beam dynamics in *PT* symmetric optical lattices, Phys. Rev. Lett. **100**, 103904 (2008).

[3] A. Guo, G. J. Salamo, D. Duchesne, R. Morandotti, M. Volatier-Ravat, V. Aimez, G. A. Siviloglou and D. N. Christodoulides, Observation of *PT*-symmetry breaking in complex optical potentials, Phys. Rev. Lett. **103**, 093902 (2009).

[4] S. Longhi, Bloch oscillations in complex crystals with *PT* symmetry, Phys. Rev. Lett. **103**, 123601 (2009).

[5] S. Longhi, Optical realization of relativistic non-Hermitian quantum mechanics, Phys. Rev. Lett. **105**, 013903 (2010).

[6] C. E. Rüter, K. G. Makris, R. El-Ganainy, D. N. Christodoulides, M. Segev and D. Kip, Observation of parity–time symmetry in optics, Nat. Phys. **6**, 192 (2010).

[7] A. Regensburger, C. Bersch, M. A. Miri, G. Onishchukov, D. N. Christodoulides and U. Peschel, Parity-time synthetic photonic lattices, Nature **488**, 167 (2012).

[8] L. Chang, X. Jiang, S. Hua, C. Yang, J. Wen, L. Jiang, G. Li, G. Wang and M. Xiao, Parity-time symmetry and variable optical isolation in active-passive-coupled microresonators, Nat. Photon. **8**, 524 (2014).

[9] L. Ge and A. D. Stone, Parity-Time Symmetry Breaking beyond One Dimension: The Role of Degeneracy, Phys. Rev. X **4**, 031011 (2014).

[10] Y. D. Chong, L. Ge, H. Cao and A. D. Stone, Coherent perfect absorbers: time-reversed lasers, Phys. Rev. Lett. **105**, 053901 (2010).

[11] L. Feng, Z. J. Wong, R.-M. Ma, Y. Wang and X. Zhang, Single-mode laser by parity-time symmetry breaking, Science **346**, 972 (2014).

[12] Y. D. Chong, L. Ge and A. D. Stone, *PT*-symmetry breaking and laser-absorber modes in optical scattering systems, Phys. Rev. Lett. **106**, 093902 (2011).

[13] H. Hodaei, M. A. Miri, M. Heinrich, D. N. Christodoulides and M. Khajavikhan, Parity-time-symmetric microring lasers, Science **346**, 975 (2014).

[14] P. A. Kalozoumis, C. V. Morfonios, F. K. Diakonos and P. Schmelcher, *PT*-symmetry breaking in waveguides with competing loss-gain pairs, Phys. Rev. A **93**, 063831 (2016).

[15] Y. Sun, W. Tan, H. Q. Li, J. Li and H. Chen, Experimental demonstration of a coherent perfect absorber with *PT* phase transition, Phys. Rev. Lett. **112**, 143903 (2014).

[16] Z. Lin, H. Ramezani, T. Eichelkraut, T. Kottos, H. Cao and D. N. Christodoulides, Unidirectional invisibility induced by *PT*-symmetric periodic structures, Phys. Rev. Lett. **106**, 213901 (2011).

[17] X. Zhu, H. Ramezani, C. Shi, J. Zhu and X. Zhang, *PT*-Symmetric Acoustics, Phys. Rev. X **4**, 031042 (2014).

[18] L. Sounas, R. Fleury and A. Alu, Unidirectional Cloaking Based on Metasurfaces with Balanced Loss and Gain, Phys. Rev. Appl. **4**, 014005 (2015).

[19] R. Fleury, D. Sounas and A. Alu, An invisible acoustic sensor based on parity-time symmetry, Nat. Commun. **6**, 5905 (2015).

[20] C. Shi, M. Dubois, Y. Chen, L. Cheng, H. Ramezani, Y. Wang and X. Zhang, Accessing the exceptional points of parity-time symmetric acoustics, Nat. Commun. **7**, 11110 (2016).

[21] L. Feng, Y. L. Xu, W. S. Fegadolli, M. H. Lu, J. E. Oliveira, V. R. Almeida, Y. F. Chen and A. Scherer, Experimental demonstration of a unidirectional reflectionless parity-time metamaterial at optical frequencies, Nat. Mater. **12**, 108 (2013).

[22] T. Liu, X. F. Zhu, F. Chen, S. J. Liang and J. Zhu, Unidirectional Wave Vector Manipulation in Two-Dimensional Space with an All Passive Acoustic Parity-Time-Symmetric Metamaterials Crystal, Phys. Rev. Lett. **120**, 124502 (2018).

[23] X.-Y. Zhu, Y.-L. Xu, Y. Zou, X.-C. Sun, C. He, M.-H. Lu, X.-P. Liu and Y.-F. Chen, Asymmetric diffraction based on a passive parity-time grating, Appl. Phys. Lett. **109**, 111101 (2016).

[24] Z. J. Wong, Y.-L. Xu, J. Kim, K. O'Brien, Y. Wang, L. Feng and X. Zhang, Lasing and anti-lasing in a single cavity, Nat. Photon. **10**, 796 (2016).

[25] Y.-L. Xu, W. S. Fegadolli, L. Gan, M.-H. Lu, X.-P. Liu, Z.-Y. Li, A. Scherer and Y.-F. Chen, Experimental realization of Bloch oscillations in a parity-time synthetic silicon photonic, Nat. Commun. **7**, 11319 (2016).

[26] V. Achilleos, Y. Auregan and V. Pagneux, Scattering by Finite Periodic *PT*-Symmetric Structures, Phys. Rev. Lett. **119**, 243904 (2017).

[27] Y. Auregan and V. Pagneux, *PT*-Symmetric Scattering in Flow Duct Acoustics, Phys. Rev. Lett. **118**, 174301 (2017).

[28] T. Goldzak, A. A. Mailybaev and N. Moiseyev, Light Stops at Exceptional Points, Phys. Rev. Lett. **120**, 013901 (2018).

[29] L. Ge and L. Feng, Contrasting eigenvalue and singular-value spectra for lasing and antilasing in a *PT*-symmetric periodic structure, Phys. Rev. A **95**, 013813 (2017).

[30] V. A. Bushuev, L. V. Dergacheva and B. I. Mantsyzov, Asymmetric pendulum effect and transparency change of *PT*-symmetric photonic crystals under dynamical Bragg diffraction beyond the paraxial approximation, Phys. Rev. A **95**, 033843 (2017).

[31] Y. Yang, H. Jia, Y. Bi, H. Zhao, J. Yang, Experimental Demonstration of an Acoustic Asymmetric Diffraction Grating Based on Passive Parity-Time-Symmetric Medium, Phys. Rev.

Appl. **12**, 034040 (2019).

[32] M. H. Lu, X. K. Liu, L. Feng, J. Li, C. P. Huang, Y. F. Chen, Y. Y. Zhu, S. N. Zhu and N. B. Ming, Extraordinary acoustic transmission through a 1D grating with very narrow apertures, Phys. Rev. Lett. **99**, 174301 (2007).

[33] F. Cai, F. Liu, Z. He and Z. Liu, High refractive-index sonic material based on periodic subwavelength structure, Appl. Phys. Lett. **91**, 203515 (2007).

[34] J. Christensen, L. Martin-Moreno and F. J. Garcia-Vidal, Theory of Resonant Acoustic Transmission through Subwavelength Apertures, Phys. Rev. Lett. **101**, 014301 (2008).

[35] F. G. Kaspar, Diffraction by Thick, Periodically Stratified Gratings with Complex Dielectric-Constant, J. Opt. Soc. Am. **63**, 37 (1973).

[36] K. Knop, Rigorous Diffraction Theory for Transmission Phase Gratings with Deep Rectangular Grooves, J. Opt. Soc. Am. **68**, 1206 (1978).

[37] M. G. Moharam and T. K. Gaylord, Rigorous Coupled-Wave Analysis of Planar-Grating Diffraction, J. Opt. Soc. Am. **71**, 811 (1981).

[38] M. G. Moharam and T. K. Gaylord, Rigorous Coupled-Wave Analysis of Grating Diffraction - E-Mode Polarization and Losses, J. Opt. Soc. Am. **73**, 451 (1983).

[39] Z. Zylberberg and E. Marom, Rigorous Coupled-Wave Analysis of Pure Reflection Gratings, J. Opt. Soc. Am. **73**, 392 (1983).

[40] M. G. Moharam and T. K. Gaylord, Rigorous coupled-wave analysis of metallic surface-relief gratings, J. Opt. Soc. Am. A **3**, 1780 (1986).

[41] N. Chateau and J. P. Hugonin, Algorithm for the Rigorous Coupled-Wave Analysis of Grating Diffraction, J. Opt. Soc. Am. A **11**, 1321 (1994).

[42] M. G. Moharam, E. B. Grann, D. A. Pommet and T. K. Gaylord, Formulation for Stable and Efficient Implementation of the Rigorous Coupled-Wave Analysis of Binary Gratings, J. Opt. Soc. Am. A **12**, 1068 (1995).

[43] C. C. Tsai and S. T. Wu, Study of broadband polarization conversion with metallic surface-relief gratings by rigorous coupled-wave analysis, J. Opt. Soc. Am. A **25**, 1339 (2008).